\documentclass[10pt,letterpaper,twocolumn,american,nofootinbib]{revtex4}
\usepackage{ae}
\usepackage{aecompl}
\usepackage[T1]{fontenc}
\usepackage[latin1]{inputenc}
\usepackage{amsmath}
\usepackage{graphicx}
\usepackage{amssymb}

\makeatletter
\usepackage{psfrag}

\usepackage{babel}
\makeatother
\begin{document}

\title{Age-dependent decay in the landscape}

\author{Sergei Winitzki}

\affiliation{Department of Physics, University of Heidelberg, Germany; on leave
from Arnold Sommerfeld Center, Department of Physics, University of
Munich, Germany}

\begin{abstract}
The picture of the {}``multiverse'' arising in diverse cosmological
scenarios involves transitions between metastable vacuum states. It
was pointed out by Krauss and Dent that the transition rates decrease
at very late times, leading to a dependence of the transition probability
between vacua on the age of each vacuum region. I investigate the
implications of this non-Markovian, age-dependent decay on the global
structure of the spacetime in landscape scenarios. I show that the
fractal dimension of the eternally inflating domain is precisely equal
to 3, instead of being slightly below 3 in scenarios with purely Markovian,
age-independent decay. I develop a complete description of a non-Markovian
landscape in terms of a nonlocal master equation. Using this description
I demonstrate by an explicit calculation that,
under some technical assumptions about the landscape,
the probabilistic predictions
of our position in the landscape are essentially unchanged, regardless
of the measure used to extract these predictions. I briefly discuss
the physical plausibility of realizing non-Markovian vacuum decay
in cosmology in view of the possible decoherence of the metastable
quantum state.
\end{abstract}
\maketitle

\section{Introduction}

According to today's accepted cosmological data, the universe is now
undergoing accelerated expansion with an approximately constant Hubble
rate $H_{\text{now}}$. However, models of string theory suggest that
this accelerating state may be merely a metastable vacuum that is
destined, after a long time, to decay via quantum tunneling into other
states with different values of $H$. The recently developed paradigm
of {}``string theory landscape''~\cite{Susskind:2003kw} involves
a very large number of metastable vacua, corresponding to local minima
of an effective potential in field space. The value of the potential
at each minimum determines the effective Hubble rate in the corresponding
vacuum. A similar scenario combining inflationary evolution and tunneling
was proposed earlier in Ref.~\cite{Garriga:1997ef} under the name
of {}``recycling universe.'' In all these scenarios, the universe
becomes a {}``multiverse,'' that is, an infinite ensemble of large,
causally disconnected spatial regions. Some of these regions contain
galaxies and stars, while other regions are undergoing inflation and
generating new vast domains of space. Each spatial domain may be in
a metastable vacuum state with a sufficiently long decay time, so
that reheating can occur and the standard cosmological evolution can
proceed before the transition to a different vacuum state.

The theory allows in principle to determine the set of possible vacua
but does not predict our position in the landscape with certainty.
After many transitions, the position of our observable patch of the
universe in the landscape becomes random. Nevertheless, one would
like to explain the present value of the cosmological constant and
possibly other observables. Therefore one attempts to calculate the
\emph{probability} of being in a vacuum of a given kind, for a {}``typical''
observer. It is notoriously difficult to formulate an unambiguous
and well-behaved measure on the set of all possible observers such
that the {}``typical'' observers are selected without bias; see
e.g.~\cite{Maor:2007wq,Garriga:2007wz} for a recent discussion and
Refs.~\cite{Aguirre:2006ak,Winitzki:2006rn,Vilenkin:2006xv,Guth:2007ng}
for reviews of the proposals of observer-based measure. 

In this paper I study a different aspect of the measure problem. All
currently proposed measures are based on the assumption that the decay
of a metastable state proceeds independently of the individual age
of that state. In other words, it is assumed that the random process
of transitions between different states in the landscape is a Markov
chain. Markovian transition probabilities are determined only by the
current state and have no memory of previous transitions. (The {}``memory''
effect due to bubble collisions~\cite{Garriga:2006hw} does not modify
transition probabilities.) Vacuum decay proceeds through bubble nucleation
and is normally described via the nucleation rate per unit 4-volume~\cite{Coleman:1980aw,Garriga:1993fh},\begin{equation}
\Gamma^{(4D)}=O(1)H^{4}\exp\left[-S_{I}-\frac{\pi}{H^{2}}\right],\label{eq:nucleation rate}\end{equation}
where $S_{I}$ is the relevant instanton action and $H$ is the Hubble
rate of the parent vacuum (I use the Planck units throughout the paper).
The transition rates between different metastable vacua can be considered
(in principle) known in a given model of the landscape. For a fixed
3-volume $V$, the probability of nucleating no bubbles after time
$t$ is exponentially small, $\propto\exp[-\Gamma^{(4D)}Vt]$.

A statistical description of evolution in the landscape can be obtained~\cite{Garriga:1997ef,Garriga:2005av}
by considering the fraction $f_{\alpha}(t)$ of the comoving volume
occupied by bubbles of type $\alpha$ at time $t$. One can approximate
the transition to a different vacuum as a series of random nucleation
events, each event resulting in an instantaneous conversion of a volume
$H_{\alpha}^{-3}$ of vacuum type $\alpha$ into the same volume of
vacuum type $\beta$. The rate of this conversion per unit time, denoted
$\kappa_{\alpha\rightarrow\beta}$, can be computed according to Eq.~(\ref{eq:nucleation rate})
with appropriate normalization factor,\begin{equation}
\kappa_{\alpha\rightarrow\beta}=O(1)H_{\alpha}\exp\left[-S_{\alpha\rightarrow\beta}-\frac{\pi}{H_{\alpha}^{2}}\right].\end{equation}
Defining for convenience $\kappa_{\alpha\rightarrow\alpha}\equiv0$,
one then writes the master equation describing the evolution of $f_{\alpha}(t)$,\begin{equation}
\frac{df_{\alpha}}{dt}=\sum_{\beta}\left(\kappa_{\beta\rightarrow\alpha}f_{\beta}-\kappa_{\alpha\rightarrow\beta}f_{\alpha}\right).\label{eq:master equation}\end{equation}
This equation can be solved with initial conditions $f_{\alpha}(0)$.
Some measure prescriptions based on the comoving distribution were
proposed in Refs.~\cite{Freivogel:2005vv,Bousso:2006ev,Vanchurin:2006xp}.

Another useful distribution is the 3-volume $V_{\alpha}(t)$ of spatial
regions within bubbles of type $\alpha$ at time $t$. The evolution
equation for $V_{\alpha}(t)$ differs from Eq.~(\ref{eq:master equation})
by the volume expansion factors,\begin{align}
\frac{dV_{\alpha}}{dt} & =3H_{\alpha}V_{\alpha}+\sum_{\beta}\left(\kappa_{\beta\rightarrow\alpha}f_{\beta}-\kappa_{\alpha\rightarrow\beta}f_{\alpha}\right)\label{eq:master equation volume}\\
 & \equiv\sum_{\beta}M_{\alpha\beta}V_{\beta},\end{align}
where the matrix $M_{\alpha\beta}$ is defined by \begin{equation}
M_{\alpha\beta}\equiv\left(3H_{\alpha}-\sum_{\lambda}\kappa_{\alpha\rightarrow\lambda}\right)\delta_{\alpha\beta}+\kappa_{\beta\rightarrow\alpha}.\label{eq:matrix def}\end{equation}
The volume-weighted master equations are used in volume-based measure
prescriptions (e.g.~Refs.~\cite{Garriga:2005av,Linde:2007nm}). 

All the existing measure prescriptions depend on the properties of
the \emph{late}-\emph{time} behavior of the distributions $f_{\alpha}(t)$
and $V_{\alpha}(t)$. The late-time asymptotics of the solutions of
Eqs.~(\ref{eq:master equation})--(\ref{eq:master equation volume})
are always exponential. For instance, the volume distribution has
the late-time asymptotics $V_{\alpha}\propto c_{\alpha}e^{\gamma t}$,
where $\gamma>0$ is the dominant eigenvalue of the matrix $M_{\alpha\beta}$.
The values of the coefficients $c_{\alpha}$ are determined by the
right eigenvector of $M_{\alpha\beta}$ corresponding to the eigenvalue
$\gamma$. 

Recently, Krauss and Dent~\cite{Krauss:2007rx} called attention
to the fact that the decay of metastable states becomes subexponential
at very late times. In typical quantum-mechanical metastable systems
in $d$-dimensional space, the probability of not decaying (the {}``survival
probability'') initially decreases exponentially as $e^{-\Gamma t}$,
where $\Gamma$ is the decay rate, but eventually starts falling off
as $t^{-d}$ after a (very long) crossover time $T\sim5\Gamma^{-1}\ln\left(E/\Gamma\right)$,
where $E$ is the energy difference between the metastable state and
the final stable state. Effectively, the tunneling rate for all transitions
between states goes to zero as $\Gamma(t)\propto t^{-1}$ after a
(state-dependent) crossover time. It is important to note that the
transition dynamics depends on the {}``age'' of the current state,
i.e.~on the time elapsed since the last transition. With this modification,
the transition process becomes a non-Markov random walk, and Eqs.~(\ref{eq:master equation})--(\ref{eq:master equation volume})
no longer apply. In particular, the late-time asymptotics of the bubble
distributions $f_{\beta}(t)$ and $V_{\beta}(t)$ are no longer purely
exponential. For this reason it is interesting to investigate the
implications of the non-Markov transitions for the measure calculations,
which depend in an essential way on the late-time behavior of $f_{\beta}(t)$
and $V_{\beta}(t)$.

In this paper I study the evolution of the landscape assuming that
the late-time asymptotic of the survival probability becomes subexponential
at a state-dependent crossover time. The main results of this first
study are as follows. I show that the fractal dimension of the inflating
domain is exactly equal to 3, while it is always slightly below 3
in Markovian models. Then I develop an explicit non-Markovian description
of the transition dynamics in terms of a master equation that is nonlocal
in time. Using that equation, I derive the late-time asymptotics of
the volume distributions $V_{\alpha}(t)$ using the proper time coordinate
$t$. The results show explicitly, within a controlled approximation,
that the volume ratios $V_{\alpha}(t)/V_{\beta}(t)$ approach a constant
at late times and are approximately the same as those computed within
the Markovian situation, except for the volume in bubbles of type
0 having the largest Hubble rate $H_{0}=\max_{\alpha}H_{\alpha}$.
The bubbles of type 0 now entirely dominate the volume of the universe
at a fixed time $t$, whereas their volume fraction was large but
finite in Markovian scenarios. These results (obtained using the proper time gauge)
applied to landscapes where
a single vacuum type has the largest Hubble rate of all available vacuum types.
I also show that the comoving volume
distributions remain essentially unchanged in the non-Markovian regime.
This suggests that the results obtained in any measure prescription
(whether volume-based or worldline-based) do not need any modification
in view of the modified late-time decay. I conclude with a brief discussion
of the viability of the non-Markovian assumption in the cosmological
context.

\section{Non-Markovian Sierpi\'nski carpet\label{sec:Non-Markovian-Sierpi}}

\begin{figure}
\begin{centering}
\psfrag{inflating}{inflating regions}\psfrag{thermalized}{thermalized regions}\includegraphics[width=3in]{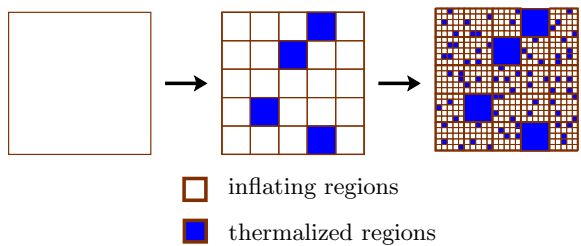}
\par\end{centering}

\caption{First steps in the construction of a random Sierpi\'nski carpet
with $N=5$ and $q=5/6$. \label{cap:square}}
\end{figure}

I begin by examining the global structure of the spacetime undergoing
non-Markovian vacuum decay. A particular version of the random Sierpi\'nski
carpet, or {}``inflation in a box,'' was considered in Ref.~\cite{Winitzki:2005ya}
as a drastically simplified toy model mimicking the global geometry
of such a spacetime. In this model, time elapses in discrete steps,
and the space is reduced to a two-dimensional square domain $0<x,y<1$,
where $x,y$ are the \emph{comoving} coordinates. The entire initial
Hubble-size domain is assumed to be initially inflating. To imitate
inflation during one time step, one subdivides the initial inflating
square into $N\times N$ equal sub-squares of size $N^{-1}\times N^{-1}$;
at the next step, each sub-square will again have the Hubble proper
size. Then one randomly marks some of the smaller squares as {}``thermalized,''
assuming that each Hubble-size inflating square continues inflation
with a probability $q$ (where $0<q<1$) and thermalizes with probability
$1-q$, independently of all other squares. The selection of thermalized
squares concludes the simulation for one timestep. At the next timestep,
the same procedure of subdivision and random thermalization is applied
to each Hubble-sized inflating square, while the {}``thermalized''
squares do not evolve any further (see Fig.~\ref{cap:square}). This
process is continued indefinitely and generates a fractal set of measure
zero consisting of points that never enter any thermalized squares
(called the {}``eternal points'' in Ref.~\cite{Winitzki:2001np}
where rigorous definitions are given). This set represents the eternally
inflating subdomain of the spacetime. Under the condition $N^{2}q>1$,
the fractal dimension of the eternally inflating domain is $\gamma=2+\ln q/\ln N>0$,
and future-eternal inflation occurs with a nonzero probability~\cite{Winitzki:2005ya}. 

As formulated, the model is Markovian since the thermalization probability
at each step is independent of the age of the inflating square. The
probability of remaining in the inflationary regime (the {}``survival
probability'') after $t$ time steps is $q^{t}=e^{-\alpha t}$, where
$\alpha\equiv\ln\frac{1}{q}$. Let us now modify this toy model by
assuming that the survival probability is given by a function $S(t)$
that interpolates between the initially exponential falloff $S(t)=e^{-\alpha t}$
for $t\ll T$ and the power-law asymptotic $S(t)\approx S_{0}t^{-p}$
for $t\gg T$, where $p$ is a fixed constant and $T$ is the crossover
time. We would like to compute the fractal dimension of the set of
eternal points in this non-Markovian model.

Let us denote by $X(t)$ the probability of the presence of at least
one eternal point within an inflating square at time $t$. The quantity
$X(t)$ can be computed explicitly, but it is sufficient for the present
purposes to obtain the asymptotic value of $X(t)$ at $t\rightarrow\infty$.
Since the thermalization probability per step goes to zero at late
times, the value of $X(t)$ approaches 1 as $t\rightarrow\infty$.
More precisely, $X(t)$ is the nonzero solution of the equation \begin{equation}
1-X(t)=p(t)+\left(1-p(t)\right)\left(1-X\right)^{N^{2}}.\end{equation}
An approximate solution of this equation for $p(t)\ll1$ is $1-X\approx p(t)$.
Since $p(t)\rightarrow0$ as $t\rightarrow\infty$, we have $X(t)\rightarrow1$.
Hence the average number of inflating squares containing at least
one eternal point at a late time $t$ is $\approx S(t)N^{2t}$, while
the linear size of each square is $N^{-1}$. So the fractal dimension
of the eternally inflating set is \begin{equation}
\gamma=2-\lim_{t\rightarrow\infty}\frac{\ln S(t)}{\ln N^{-t}}=2+\lim_{t\rightarrow\infty}\frac{\ln\left(S_{0}t^{-p}\right)}{t\ln N}=2.\end{equation}
It can be shown that the eternally inflating domain consists of an
infinite merged cluster when it is formed as a random Sierpi\'nski
carpet with fractal dimension 2. It is important to note that the
eternal set still has measure zero because every comoving point will
reach thermalization with probability 1.

By analogy, one can investigate the eternally inflating domain in
a three-dimensional space and conclude that its fractal dimension
is 3. A quick argument leading to this conclusion consists of estimating
the growth of the 3-volume of the inflating domain as $V(t)\propto e^{3Ht}t^{-p}$
in the regime of power-law decay at late times. A domain growing as
$V(t)\propto e^{\gamma Ht}$ is interpreted as a lacunary fractal
with dimension $\gamma$~\cite{Aryal:1987vn,Winitzki:2005ya}, regardless
of subexponential corrections. Therefore, the fractal dimension of
the inflating domain is always equal to 3 in the non-Markovian case.
This is only a small correction to the results obtained in typical
scenarios of eternal inflation where the fractal dimension is very
slightly below 3 (see, for instance, Refs.~\cite{Aryal:1987vn,Linde:1993xx}).
Therefore, the global geometry of the spacetime is not significantly
modified in these scenarios even if the late-time decay is subexponential.

\section{Evolution in a non-Markovian landscape}

The next issue is whether the results of applying the various landscape
measure proposals are modified when non-Markovian decay is assumed.
In this section I derive the suitably modified versions of Eqs.~(\ref{eq:master equation})--(\ref{eq:master equation volume})
and obtain their late-time asymptotics. Since all the different measure
proposals require computing the late-time behavior of these same evolution
equations, the results of the present calculation will be equally
relevant to every measure proposal.

To describe the evolution of spacetime in a landscape scenario with
non-Markovian transitions, one needs to specify the transition rate
$\Gamma_{\alpha\rightarrow\beta}(t)$ between vacua $\alpha$ and
$\beta$ as a function of the age $t$ of the parent vacuum $\alpha$.
The precise form of $\Gamma_{\alpha\rightarrow\beta}(t)$ will be
model-dependent except for the properties $\Gamma_{\alpha\rightarrow\beta}(t)\approx\kappa_{\alpha\rightarrow\beta}=\mbox{const}$
for $t<T_{\alpha\rightarrow\beta}$, where $T_{\alpha\rightarrow\beta}$
is the crossover time, and $\Gamma_{\alpha\rightarrow\beta}(t)\rightarrow0$
for $t\gg T_{\alpha\rightarrow\beta}$. For simplicity I will assume
below that the crossover time $T_{\alpha\rightarrow\beta}\equiv T$
is independent of $\alpha$ and $\beta$. Without this technical assumption,
the analysis will be more complicated without yielding significantly
different results. If transitions $\alpha\rightarrow\beta$ have different
crossover times $T_{\alpha\rightarrow\beta}$, the results of the
present analysis will be approximately applicable at sufficiently
late times $t$ such that $t\gg T\equiv\max_{\alpha,\beta}T_{\alpha\rightarrow\beta}$.

Since transition probabilities depend on the age, it is not sufficient
to consider the probability distributions $f_{\beta}(t)$ and $V_{\beta}(t)$
mentioned above. One needs to introduce more detailed distributions
that include information about the times of the previous transitions.

\subsection{Volume distributions}

I first consider the volume distribution. Assume for convenience that
there is a single initial bubble of type $\alpha_{0}$ formed at time
$t=0$ with unit volume, and that we are interested in describing
only the evolution of the interior of the initial bubble and any bubbles
nucleated in it. (The case of several initial bubbles is a straightforward
extension.) Let $V_{\alpha}(t_{0},t)dt_{0}$ denote the volume at
time $t$ of bubbles of type $\alpha$ that were formed at an earlier
time between $t_{0}$ and $t_{0}+dt_{0}$. By definition, we set $V_{\alpha}(t_{0},t)=0$
for $t_{0}>t$. The volume remaining from the initial bubble could
be included in $V_{\alpha_{0}}(t_{0},t)$ as a contribution of the
form $\delta(t_{0})V_{\alpha_{0}}^{(0)}(t)$, but it is techically
more convenient to \emph{exclude} the initial bubble from $V_{\alpha}(t_{0},t)$
and to account for its volume $V_{\alpha_{0}}^{(0)}(t)$ separately.
The quantity $V_{\alpha_{0}}^{(0)}(t)$ represents the proper volume
that remains from the initial bubble and has not decayed by time $t$.
The volume $V_{\alpha_{0}}^{(0)}(t)$ of the initial bubble grows
with the rate $3H_{\alpha_{0}}$ and decreases due to nucleation of
other bubbles:\begin{equation}
\frac{dV_{\alpha_{0}}^{(0)}(t)}{dt}=3H_{\alpha_{0}}V_{\alpha_{0}}^{(0)}(t)-\sum_{\beta}\Gamma_{\alpha_{0}\rightarrow\beta}(t)V_{\alpha_{0}}^{(0)}(t).\label{eq:orig bubble V}\end{equation}
Integrating Eq.~(\ref{eq:orig bubble V}) with the initial condition
$V_{\alpha_{0}}^{(0)}(0)=1$, we find\begin{align}
V_{\alpha_{0}}^{(0)}(t) & =\exp\left[3H_{\alpha_{0}}t\right]S_{\alpha_{0}}(t),\\
S_{\alpha}(t) & \equiv\exp\left[-\int_{0}^{t}\sum_{\beta}\Gamma_{\alpha\rightarrow\beta}(t')dt'\right].\end{align}
The auxiliary function $S_{\alpha}(t)$ is the survival probability
of a bubble of type $\alpha$ and age $t$. 

The evolution equation for $V_{\alpha}(t_{0},t)$ accounts for the
growth of volume at rate $3H_{\alpha}$, age-dependent decay into
bubbles of different kinds, and age-dependent nucleation of zero-age
bubbles of kind $\alpha$ from other bubbles (including the original
bubble): \begin{align}
\frac{\partial V_{\alpha}(t_{0},t)}{\partial t} & =3H_{\alpha}V_{\alpha}(t_{0},t)-\sum_{\beta}\Gamma_{\alpha\rightarrow\beta}(t-t_{0})V_{\alpha}(t_{0},t)\nonumber \\
 & +\delta(t-t_{0})\negmedspace\int_{0}^{t}\negmedspace\negmedspace d\tilde{t}_{0}\sum_{\beta}\Gamma_{\beta\rightarrow\alpha}(t-\tilde{t}_{0})V_{\beta}(\tilde{t}_{0},t)\nonumber \\
 & +\delta(t-t_{0})\Gamma_{\alpha_{0}\rightarrow\alpha}(t)V_{\alpha_{0}}^{(0)}(t).\label{eq:evolution V alpha}\end{align}
The factors $\delta(t-t_{0})$ account for the fact that bubbles nucleated
at time $t$ have zero age at that time and therefore contribute to
the distribution $V_{\alpha}(t_{0},t$) only at $t_{0}=t$.

Noting that $V_{\alpha}(t_{0},t)$ with $t\neq t_{0}$ is decoupled
from other $V_{\beta}(t_{0},t)$, we have (for $t>t_{0}$) \begin{equation}
V_{\alpha}(t_{0},t)=V_{\alpha}(t_{0},t_{0})\exp\left[3H_{\alpha}\left(t-t_{0}\right)\right]S_{\alpha}(t-t_{0}).\label{eq:V alpha solution 1}\end{equation}
It remains to determine the function $V_{\alpha}(t_{0},t_{0})\equiv U_{\alpha}(t_{0})$.
Integrating Eq.~(\ref{eq:evolution V alpha}) in $t$ over an infinitesimal
interval around $t=t_{0}$ and using Eq.~(\ref{eq:V alpha solution 1})
and the condition $V_{\alpha}(t_{0},t)=0$ for $t<t_{0}$, we obtain
a closed system of integral equations for $U_{\alpha}(t)$, \begin{align}
U_{\alpha}(t) & =\negmedspace\sum_{\beta}\negmedspace\int_{0}^{t}\negmedspace\negmedspace d\tilde{t}_{0}U_{\beta}(\tilde{t}_{0})e^{3H_{\beta}(t-\tilde{t}_{0})}S_{\beta}(t-\tilde{t}_{0})\Gamma_{\beta\rightarrow\alpha}(t-\tilde{t}_{0})\label{eq:integral equ V alpha integral}\\
 & +\Gamma_{\alpha_{0}\rightarrow\alpha}(t)e^{3H_{\alpha_{0}}t}S_{\alpha_{0}}(t).\label{eq:integral equation V alpha}\end{align}
It remains to determine the asymptotic behavior of the functions $U_{\alpha}(t)$.

\subsubsection{Markovian regime}

I first consider times $t$ before the crossover time scale, $0<t<T$.
At these times, the behavior of the system is (approximately) Markovian,
and one expects to recover the standard equations~(\ref{eq:master equation volume}).
In Eq.~(\ref{eq:integral equation V alpha}) we may approximately
set \begin{align}
\Gamma_{\beta\rightarrow\alpha}(t-\tilde{t}_{0}) & \approx\kappa_{\beta\rightarrow\alpha}=\mbox{const},\\
S_{\beta}(t-\tilde{t}_{0}) & \approx\exp\left[-\left(t-\tilde{t}_{0}\right)\Gamma_{\beta}\right],\end{align}
where we denoted by $\Gamma_{\beta}\equiv\sum_{\alpha}\kappa_{\beta\rightarrow\alpha}$
the total decay rate of the vacuum type $\beta$ in the Markovian
regime. Then Eq.~(\ref{eq:integral equation V alpha}) is rewritten
as \begin{align}
U_{\alpha}(t) & \approx\negmedspace\sum_{\beta}\kappa_{\beta\rightarrow\alpha}\negmedspace\int_{0}^{t}\negmedspace\negmedspace d\tilde{t}_{0}U_{\beta}(\tilde{t}_{0})e^{\left(3H_{\beta}-\Gamma_{\beta}\right)(t-\tilde{t}_{0})}\nonumber \\
 & +\Gamma_{\alpha_{0}\rightarrow\alpha}e^{(3H_{\alpha_{0}}-\Gamma_{\alpha_{0}})t}.\label{eq:integral equation V alpha 1}\end{align}
 Although this system of equations appears to be nonlocal in time,
it can be reduced explicitly to a Markovian system. We pass to new
variables\begin{equation}
V_{\alpha}(t)\equiv\negmedspace\int_{0}^{t}\negmedspace\negmedspace d\tilde{t}_{0}U_{\alpha}(\tilde{t}_{0})e^{\left(3H_{\alpha}-\Gamma_{\alpha}\right)(t-\tilde{t}_{0})}+\delta_{\alpha\alpha_{0}}e^{(3H_{\alpha_{0}}-\Gamma_{\alpha_{0}})t}.\end{equation}
The quantity $V_{\alpha}(t)$ represents the total volume inside bubbles
of type $\alpha$ at time $t$ integrated over the bubble ages and
also including the volume of the initial bubble. The variables $U_{\alpha}(t)$
are expressed through $V_{\alpha}(t)$ as\begin{align}
U_{\alpha}(t) & =e^{\left(3H_{\alpha}-\Gamma_{\alpha}\right)t}\partial_{t}\left[e^{-\left(3H_{\alpha}-\Gamma_{\alpha}\right)t}V_{\alpha}(t)\right]\nonumber \\
 & =\dot{V}_{\alpha}-\left(3H_{\alpha}-\Gamma_{\alpha}\right)V_{\alpha}.\end{align}
Hence the volumes $V_{\alpha}(t)$ satisfy the differential equation
\begin{equation}
\dot{V}_{\alpha}=\left(3H_{\alpha}-\Gamma_{\alpha}\right)V_{\alpha}+\sum_{\beta}\kappa_{\beta\rightarrow\alpha}V_{\beta}=\sum_{\beta}M_{\alpha\beta}V_{\beta}\end{equation}
with the initial condition $V_{\alpha}(0)=\delta_{\alpha\alpha_{0}}$,
which is equivalent to Eqs.~(\ref{eq:master equation volume}); the
matrix $M_{\alpha\beta}$ is defined by Eq.~(\ref{eq:matrix def}).
The late-time asymptotic of solutions is exponential, \begin{equation}
V_{\alpha}(t)=c_{\alpha}e^{\gamma t},\label{eq:Markovian U}\end{equation}
where $\gamma$ is the largest eigenvalue of the matrix $M_{\alpha\beta}$.
It is important to note that $\gamma>3H_{\beta}-\Gamma_{\beta}$ for
all $\beta$.

The eigenvalue $\gamma$ and the corresponding eigenvectors of $M_{\alpha\beta}$
can be estimated explicitly under some technical assumptions. To be
specific, let us denote by $H_{0}$ and $H_{1}$ the first and the
second largest values among all the $H_{\alpha}$, and let us assume
that the nucleation rates are small, \begin{equation}
\kappa_{\alpha\rightarrow\beta}\ll H_{0}-H_{1}\quad\mbox{for all }\alpha,\beta.\label{eq:assumption on kappa}\end{equation}
Since the nucleation rates are typically exponentially small, one
can disregard terms of higher order in $\kappa_{\alpha\rightarrow\beta}$.
Then the matrix $M_{\alpha\beta}$ can be represented as a diagonal
matrix $\delta_{\alpha\beta}\left(3H_{\alpha}-\Gamma_{\alpha}\right)$
with a small perturbation of order $\kappa_{\alpha\rightarrow\beta}$,
and the dominant eigenvalue is found by standard perturbation theory
as \begin{equation}
\gamma=3H_{0}-\Gamma_{0}+\sum_{\alpha\neq0}\frac{\kappa_{\alpha\rightarrow0}\kappa_{0\rightarrow\alpha}}{3H_{0}-3H_{\alpha}}+O(\kappa_{\alpha\rightarrow\beta}^{3}).\label{eq:gamma approx}\end{equation}
The second-order term in $\gamma$ will play a role below. 

The coefficients $c_{\alpha}$ in Eq.~(\ref{eq:Markovian U}) are
proportional to the components of the (right) dominant eigenvector
$r_{\alpha0}$ of $M_{\alpha\beta}$, so that $c_{\alpha}/c_{\beta}=r_{\alpha0}/r_{\beta0}$.
The ratios of components of the eigenvector $r_{\alpha0}$ can be
found approximately as \begin{equation}
\frac{r_{\alpha0}}{r_{00}}=\frac{\kappa_{0\rightarrow\alpha}}{3H_{0}-3H_{\alpha}}+O(\kappa_{\alpha\rightarrow\beta}^{2}),\quad\alpha\neq0.\end{equation}
It is useful to compute also the absolute normalization of the coefficients
$c_{\alpha}$, which will yield an explicit late-time asymptotic $V_{\alpha}(t)=c_{\alpha}e^{\gamma t}$
as a function of the initial conditions $V_{\alpha}(0)=\delta_{\alpha\alpha_{0}}$.
The time-dependent solution $V_{\alpha}(t)$ can be decomposed as\begin{equation}
V_{\alpha}(t)=\sum_{n}v_{n}r_{\alpha n}e^{\gamma_{n}t},\end{equation}
where $\gamma_{0},\gamma_{1},...$ and $r_{\alpha0}$, $r_{\alpha1}$,
...~are the eigenvalues and the corresponding (right) eigenvectors
of $M_{\alpha\beta}$. The late-time behavior of $V_{\alpha}(t)$
is dominated by $e^{\gamma_{0}t}$, where $\gamma_{0}\equiv\gamma$
is the largest eigenvalue. 

The coefficients $v_{n}$ are found by decomposing the initial condition
vector $V_{\alpha}(0)$ in the basis $\left\{ r_{\alpha n}\right\} $,\begin{equation}
V_{\alpha}(0)=\sum_{n}v_{n}r_{\alpha n}.\label{eq:V alpha 0 through v}\end{equation}
The coefficients $v_{n}$ are computed as the products of the \emph{left}
eigenvectors $l_{\alpha n}$, $n=0,1,...$ of $M_{\alpha\beta}$ with
the initial condition vector $V_{\alpha}(0)\equiv\delta_{\alpha\alpha_{0}}$,\begin{equation}
v_{n}=\sum_{\alpha}l_{\alpha n}V_{\alpha}(0)=l_{\alpha_{0}n},\end{equation}
where we assumed that the dual bases $\left\{ l_{\alpha n}\right\} $
and $\left\{ r_{\alpha n}\right\} $ are normalized, \begin{equation}
\sum_{\alpha}l_{\alpha m}r_{\alpha n}=\delta_{mn}.\end{equation}
 We are interested only in the coefficients $V_{\alpha0}$ corresponding
to the dominant eigenvalue $\gamma\equiv\gamma_{0}$, so $v_{0}=l_{\alpha_{0}0}$.
The vector $l_{\alpha0}$ is determined perturbatively under the assumption~(\ref{eq:assumption on kappa})
through the ratios\begin{align}
\frac{l_{\alpha0}}{l_{00}} & =\frac{\kappa_{\alpha\rightarrow0}}{3H_{0}-3H_{\alpha}}+O(\kappa_{\alpha\rightarrow\beta}^{2}),\quad\alpha\neq0.\end{align}
Hence, a suitable normalization of the eigenvectors is\begin{align}
r_{00}=1,\quad r_{\alpha0} & =\frac{\kappa_{0\rightarrow\alpha}}{3H_{0}-3H_{\alpha}}+O(\kappa_{\alpha\rightarrow\beta}^{2}),\quad\alpha\neq0;\label{eq:r alpha n ans}\\
l_{00}=1,\quad l_{\alpha0} & =\frac{\kappa_{\alpha\rightarrow0}}{3H_{0}-3H_{\alpha}}+O(\kappa_{\alpha\rightarrow\beta}^{2}),\quad\alpha\neq0.\label{eq:l alpha n ans}\end{align}
Now we may compute explicitly \begin{equation}
c_{\alpha}=v_{0}r_{\alpha0}=r_{\alpha0}l_{\alpha_{0}0}.\end{equation}
The full solution $U_{\alpha}(t)$ can be written as

\begin{equation}
U_{\alpha}(t)=\sum_{n}l_{\alpha_{0}n}r_{\alpha n}\left(\gamma_{n}-3H_{\alpha}+\Gamma_{\alpha}\right)e^{\gamma_{n}t},\quad t<T.\label{eq:U alpha ans 2}\end{equation}
The late-time (but still Markovian) behavior of $U_{\alpha}(t)$ is\begin{equation}
U_{\alpha}(t)\approx r_{\alpha0}l_{\alpha_{0}0}\left(\gamma-3H_{\alpha}+\Gamma_{\alpha}\right)e^{\gamma t}.\end{equation}

Although the absolute values of the coefficients $c_{\alpha}$ depend
on the initial conditions, the ratios $c_{\alpha}/c_{\beta}$ do not.
This is the standard property of Markovian models: the late-time asymptotics
do not depend on the initial conditions.

\subsubsection{Non-Markovian regime}

Having computed the early-time behavior of $U_{\alpha}(t)$, I now
consider the asymptotics of $U_{\alpha}(t)$ at late times $t$ for
which the survival probabilities $S_{\alpha}(t)$ are subexponential.
Since the decay rate is the logarithmic derivative of the survival
probability, it follows that $\Gamma_{\alpha\rightarrow\beta}(t)\propto t^{-1}$
at those times. To simplify calculations, I assume that \begin{equation}
S_{\alpha}(t)\Gamma_{\alpha\rightarrow\beta}=R_{\alpha}(t)\kappa_{\alpha\rightarrow\beta}\quad\mbox{for all }\beta,\end{equation}
where the function $R_{\alpha}(t)$ describes the transition from
the Markovian to the non-Markovian regime as\begin{equation}
R_{\alpha}(t)=\begin{cases}
\exp\left[-\Gamma_{\alpha}t\right], & t<T;\\
\exp\left[-\Gamma_{\alpha}T\right]\left(\frac{T}{t}\right)^{p_{\alpha}}, & t>T,\end{cases}\end{equation}
where $T$ is the crossover time and $p_{\alpha}>0$ are constants
of order $1$. (For the cited examples of subexponential decay with
$S_{\alpha}(t)\propto t^{-3}$ one will have to set $p_{\alpha}=4$.)
The assumption of a common time profile $R_{\alpha}(t)$ and a common
crossover time $T$, independent of the vacuum type $\alpha$ and
of the decay channel $\alpha\rightarrow\beta$, may be insufficiently
precise in some scenarios. Here I employ this technical assumption
as a first step towards a more complete calculation.

Let us first determine the ansatz for the asymptotics of $U_{\alpha}(t)$
by examining Eq.~(\ref{eq:integral equation V alpha}). Since $U_{\alpha}(t)$
receives contributions from all the subexponentially decaying states
$\beta\neq\alpha$ according to Eq.~(\ref{eq:integral equation V alpha}),
the late-time asymptotics of $U_{\alpha}(t)$ must grow at least as
fast as the fastest-growing function among $e^{3H_{\alpha}t}R_{\alpha}(t)$
for all $\alpha$. Hence, the exponential part of the asymptotic is
$U_{\alpha}(t)\propto e^{\tilde{\gamma}t}$, where $\tilde{\gamma}$
is not less than $3H_{0}$ and $H_{0}$ is the largest available value
among $H_{\alpha}$. However, the function $U_{\alpha}(t)$ cannot
grow \emph{faster} than $e^{3H_{0}t}$, i.e.~as $e^{\tilde{\gamma}t}$
with $\tilde{\gamma}>3H_{0}$, because in that case the integral in
line~(\ref{eq:integral equ V alpha integral}) is dominated by $\tilde{t}_{0}\approx t$
(recently nucleated bubbles) where the survival probabilities $S_{\beta}(t-\tilde{t}_{0})$
are Markovian. So the Markovian calculation leading to Eq.~(\ref{eq:integral equation V alpha 1})
still holds and yields the contradictory result $\tilde{\gamma}=\gamma\approx3H_{0}-\Gamma_{0}<3H_{0}$.
Hence, $\tilde{\gamma}=3H_{0}$. We need to allow for the possibility
that $U_{\alpha}(t)$ contains also a subexponential asymptotic, $U_{\alpha}(t)\propto e^{3H_{0}t}Q(t)$,
where $Q(t)$ is a subexponential function decaying not faster than
$R_{\alpha}(t)$ at $t\gg T$. (Below I will show that $Q(t)\propto R_{0}(t)$,
but at this point the behavior of $Q(t)$ is not yet determined.)
Thus, the late-time asymptotics of $U_{\alpha}(t)$ are of the form
\begin{equation}
U_{\alpha}(t)\approx q_{\alpha}e^{3H_{0}t}Q(t),\quad t>T,\label{eq:U alpha ansatz}\end{equation}
while the Markovian behavior was determined above in Eq.~(\ref{eq:U alpha ans 2}).
The task at hand is to determine the coefficients $q_{\alpha}$ and
the function $Q(t)$ for the non-Markovian asymptotics~(\ref{eq:U alpha ansatz}).

Let us define the auxiliary quantities\begin{align}
W_{\alpha}(t) & \equiv\negmedspace\int_{0}^{t}\negmedspace dt_{0}U_{\alpha}(t_{0})e^{3H_{\alpha}\left(t-t_{0}\right)}R_{\alpha}(t-t_{0})\nonumber \\
 & \quad+\delta_{\alpha\alpha_{0}}e^{3H_{\alpha}t}R_{\alpha}(t),\label{eq:W alpha def}\end{align}
so that Eq.~(\ref{eq:integral equation V alpha}) becomes\begin{equation}
U_{\alpha}(t)=\sum_{\beta}\kappa_{\beta\rightarrow\alpha}W_{\beta}(t).\label{eq:U alpha short equ}\end{equation}
We will first determine the asymptotics of the quantities $W_{\alpha}(t)$
for $t\gg T$.

The definition of $W_{\alpha}(t)$ involves an integral over $t_{0}$
that needs to be estimated. It is convenient to estimate it separately
for $\alpha\neq0$ and $\alpha=0$. For $\alpha\neq0$, the function
$U_{\alpha}(t_{0})$ grows as $e^{\gamma t_{0}}$ until $t_{0}=T$;
subsequently $U_{\alpha}(t_{0})$ grows even faster, as $e^{3H_{0}t}$.
This function is multiplied by a decay factor $e^{-3H_{\alpha}t_{0}}R_{\alpha}(t-t_{0})$
that never compensates the growth of $U_{\alpha}(t_{0})$ if $\alpha\neq0$
because $\gamma-3H_{\alpha}\gg\Gamma_{\alpha}$ for $\alpha\neq0$.
Therefore, the integral over $t_{0}$ is dominated by the contribution
near the upper limit $t_{0}\approx t$ where $R_{\alpha}(t-t_{0})$
is Markovian while $U_{\alpha}(t)\propto e^{3H_{0}t}$. One obtains
the asymptotic estimate \begin{equation}
W_{\alpha}(t)\approx\frac{q_{\alpha}Q(t)e^{3H_{0}t}}{3H_{0}-3H_{\alpha}+\Gamma_{\alpha}},\quad\alpha\neq0,\label{eq:W alpha estimate}\end{equation}
where the term $\propto e^{3H_{\alpha}t}R_{\alpha}(t)$ can be disregarded
since it is exponentially smaller at late times.

Estimating the quantity $W_{0}(t)$ requires somewhat more work. One
needs to split the integral in the definition of $W_{\alpha}(t)$
into three subintervals $\left[0,T\right]$, $\left[T,t-T\right]$,
and $\left[t-T,t\right]$ where different factors in the integrand
have either Markovian or non-Markovian behavior. These three integrals
are estimated as follows. The first integral, \begin{equation}
\int_{0}^{T}\negmedspace dt_{0}U_{0}(t_{0})e^{3H_{0}\left(t-t_{0}\right)}R_{0}(t-t_{0}),\end{equation}
is dominated by the contribution of $t_{0}\approx0$ because $U_{0}(t_{0})$
in the Markovian regime grows as $e^{\gamma t_{0}}$, while $\gamma<3H_{0}$.
Using Eq.~(\ref{eq:U alpha ans 2}), we find\begin{align}
 & \int_{0}^{T}\negmedspace dt_{0}U_{0}(t_{0})e^{3H_{0}\left(t-t_{0}\right)}R_{0}(t-t_{0})\nonumber \\
 & \approx e^{3H_{0}t}R(t)\negmedspace\sum_{n}v_{n}r_{0n}\frac{\gamma_{n}-3H_{0}+\Gamma_{0}}{3H_{0}-\gamma_{n}}.\label{eq:term 1 sum}\end{align}
The sum in the last line can be estimated without actually computing
all the eigenvectors $r_{0n}$ by noting that $3H_{0}-\gamma_{n}\gg\Gamma_{0}$
for $n\neq0$, and thus the factor \begin{equation}
\frac{\gamma_{n}-3H_{0}+\Gamma_{0}}{3H_{0}-\gamma_{n}}\approx-1+O(\kappa_{\alpha\rightarrow\beta}),\quad n\neq0.\end{equation}
For $n=0$ this factor is negligible,\begin{equation}
\frac{\gamma-3H_{0}+\Gamma_{0}}{\Gamma_{0}}=O(H_{0}^{-1}\kappa_{\alpha\rightarrow\beta}),\end{equation}
where we used Eq.~(\ref{eq:gamma approx}). By splitting off the
$n=0$ term from the sum in Eq.~(\ref{eq:term 1 sum}), one now obtains\begin{equation}
\sum_{n}v_{n}r_{0n}\frac{\gamma_{n}-3H_{0}+\Gamma_{0}}{3H_{0}-\gamma_{n}}\approx-\negmedspace\sum_{n\neq0}v_{n}r_{0n}.\end{equation}
The last sum can be evaluated using Eq.~(\ref{eq:V alpha 0 through v}),\begin{equation}
V_{0}(0)=\sum_{n}v_{n}r_{0n}=v_{0}r_{00}+\negmedspace\sum_{n\neq0}v_{n}r_{0n}=\delta_{0\alpha_{0}},\end{equation}
and we find\begin{align}
\sum_{n}v_{n}r_{0n}\frac{\gamma_{n}-3H_{0}+\Gamma_{0}}{3H_{0}-\gamma_{n}} & \approx v_{0}-\delta_{0\alpha_{0}}.\end{align}
Hence, the expression~(\ref{eq:term 1 sum}) is estimated as\begin{equation}
e^{3H_{0}t}R_{0}(t)\left(v_{0}-\delta_{0\alpha_{0}}\right).\end{equation}

The integral over the second interval,\begin{equation}
\int_{T}^{t-T}\negmedspace dt_{0}U_{0}(t_{0})e^{3H_{0}\left(t-t_{0}\right)}R_{0}(t-t_{0}),\end{equation}
involves both $U_{0}(t_{0})$ and $R_{0}(t-t_{0})$ in the non-Markovian
regime. We find\begin{align}
\int_{T}^{t-T} & \negmedspace dt_{0}U_{0}(t_{0})e^{3H_{0}\left(t-t_{0}\right)}R_{0}(t-t_{0})\nonumber \\
 & \approx q_{0}e^{3H_{0}t}\int_{T}^{t-T}\negmedspace dt_{0}Q(t_{0})R_{0}(t-t_{0}).\label{eq:2nd term integral}\end{align}
Since both $R_{0}(t)$ and $Q(t)$ are decaying functions, we may
estimate the integral in Eq.~(\ref{eq:2nd term integral}) as the
sum of the contributions from intervals of order $T$ at the two ends
$t_{0}=T$ and $t_{0}=t-T$,\begin{equation}
q_{0}e^{3H_{0}t}\left[R_{0}(t)Q(T)\, O(T)+Q(t)R_{0}(T)\, O(T)\right].\end{equation}
This precision is sufficient since these terms will not play a significant
role in the final result.

The integral over the third interval involves the Markovian $R_{0}(t-t_{0})$
and is dominated by $t_{0}\approx t$, \begin{equation}
\int_{t-T}^{t}\negmedspace dt_{0}U_{0}(t_{0})e^{3H_{0}\left(t-t_{0}\right)}R_{0}(t-t_{0})\approx\frac{q_{0}}{\Gamma_{0}}Q(t)e^{3H_{0}t},\end{equation}
where we disregarded $e^{-\Gamma_{0}T}\ll1$. (Note that $\Gamma_{0}T\gg1$.)

Putting together the contributions of the three intervals as well
as the last term in Eq.~(\ref{eq:W alpha def}), we obtain\begin{align}
W_{0}(t) & \approx e^{3H_{0}t}R_{0}(t)v_{0}+q_{0}e^{3H_{0}t}\Gamma_{0}^{-1}Q(t)\nonumber \\
 & +q_{0}e^{3H_{0}t}R_{0}(t)Q(T)\, O(T),\label{eq:W 0 estimate}\end{align}
where we disregarded \begin{equation}
Q(t)R_{0}(T)\, O(T)\ll q_{0}e^{3H_{0}t}\Gamma_{0}^{-1}Q(t)\end{equation}
because \begin{equation}
R_{0}(T)O(\Gamma_{0}T)=e^{-\Gamma_{0}T}O(\Gamma_{0}T)\ll1.\end{equation}

Finally, we substitute the ansatz~(\ref{eq:U alpha ansatz})
and the estimates~(\ref{eq:W alpha estimate}), (\ref{eq:W 0 estimate})
into Eqs.~(\ref{eq:U alpha short equ}) for $U_{\alpha}(t)$. In
the limit $t\gg T$, we may divide through by the factor $e^{3H_{0}t}Q(t)$
and obtain a system of equations for $q_{\alpha}$ and $Q(t)$,\begin{align}
q_{0}= & \sum_{\beta}\kappa_{\beta\rightarrow0}\frac{q_{\beta}}{3H_{0}-3H_{\beta}+\Gamma_{\beta}},\label{eq:q 0 equ 2}\\
q_{\alpha}= & \sum_{\beta}\kappa_{\beta\rightarrow\alpha}\frac{q_{\beta}}{3H_{0}-3H_{\beta}+\Gamma_{\beta}}\nonumber \\
 & +\kappa_{0\rightarrow\alpha}\left[v_{0}+q_{0}Q(T)O(T)\right]\lim_{t\rightarrow\infty}\frac{R_{0}(t)}{Q(t)},\quad\alpha\neq0.\label{eq:q alpha equ 2}\end{align}
This is an inhomogeneous linear system for $\left\{ q_{\alpha}\right\} $. 

Let us consider the possible values of $\lim_{t\rightarrow\infty}R_{0}(t)/Q(t)$
that show whether $Q(t)$ is asymptotically dominant over $R_{0}(t)$.
Since $Q(t)$ in any case does not decay faster than $R_{0}(t)$,
there are only two possibilities: either the limit is zero or it is
nonzero. I will now show that this limit must be nonzero. 

If $\lim_{t\rightarrow\infty}R_{0}(t)/Q(t)=0$, we rewrite Eqs.~(\ref{eq:q 0 equ 2})--(\ref{eq:q alpha equ 2})
as\begin{equation}
q_{\alpha}=\sum_{\beta}\kappa_{\beta\rightarrow\alpha}\frac{q_{\beta}}{3H_{0}-3H_{\beta}+\Gamma_{\beta}},\quad\alpha=0,1,...\label{eq:q alpha incorrect system}\end{equation}
Passing to auxiliary variables\begin{equation}
s_{\alpha}\equiv\frac{q_{\alpha}}{3H_{0}-3H_{\alpha}+\Gamma_{\alpha}},\label{eq:s alpha def}\end{equation}
we find\begin{equation}
3H_{0}s_{\alpha}=\sum_{\beta}M_{\alpha\beta}s_{\beta}.\end{equation}
Since the largest eigenvalue of $M_{\alpha\beta}$ is $\gamma<3H_{0}$,
it follows that $3H_{0}$ is not an eigenvalue of $M_{\alpha\beta}$.
Hence, the only solution of the homogeneous system~(\ref{eq:q alpha incorrect system})
is $q_{\alpha}=0$. This contradicts the assumption that $q_{\alpha}e^{3H_{0}t}Q(t)$
is the leading asymptotic of $U_{\alpha}(t)$. Therefore, $Q(t)$
decays exactly as $R_{0}(t)$ at late times.

Since Eq.~(\ref{eq:q alpha equ 2}) depends only on the ratio $Q(T)/Q(t)$,
the normalization of the $Q(t)$ could then be adjusted such that
$\lim_{t\rightarrow\infty}R_{0}(t)/Q(t)=1$. The value $Q(T)$ is
of order $e^{-\Gamma_{0}T}$ due to the continuity requirement \begin{equation}
U_{\alpha}(T)\approx q_{\alpha}e^{3H_{0}T}Q(T)\approx c_{\alpha}e^{3\gamma T}.\end{equation}
Therefore, the term $q_{0}Q(T)O(T)$ in Eq.~(\ref{eq:q alpha equ 2})
is exponentially small and can be neglected. We note, however, that
its magnitude depends on the initial conditions through the coefficient
$c_{\alpha}\sim O(\Gamma_{0})$, which introduces, strictly speaking,
an exponentially small dependence on initial conditions, of order
$O(\Gamma_{0}T)e^{-\Gamma_{0}T}$. 

Finally, we rewrite Eqs.~(\ref{eq:q 0 equ 2})--(\ref{eq:q alpha equ 2})
through the variables $s_{\alpha}$ as\begin{equation}
3H_{0}s_{\alpha}-\sum_{\beta}M_{\alpha\beta}s_{\beta}=v_{0}\kappa_{0\rightarrow\alpha}.\end{equation}
This is an inhomogeneous system of equations with a nondegenerate
matrix, and so the solution is unique. It follows that all $s_{\alpha}$
are of order $v_{0}\kappa_{\alpha\rightarrow\beta}$, so an approximate
expression for the solution is readily found as\begin{align}
s_{0} & \approx\frac{v_{0}}{\Gamma_{0}}\sum_{\beta}\frac{\kappa_{\beta\rightarrow0}\kappa_{0\rightarrow\beta}}{3H_{0}-3H_{\beta}},\\
s_{\alpha} & \approx\frac{v_{0}\kappa_{0\rightarrow\alpha}}{3H_{0}-3H_{\alpha}},\quad\alpha\neq0.\end{align}
The corresponding values of $q_{\alpha}$ (neglecting higher orders
of $\kappa_{\alpha\rightarrow\beta}$) are\begin{align}
q_{0} & \approx v_{0}\sum_{\beta}\frac{\kappa_{\beta\rightarrow0}\kappa_{0\rightarrow\beta}}{3H_{0}-3H_{\beta}},\\
q_{\alpha} & \approx v_{0}\kappa_{0\rightarrow\alpha},\quad\alpha\neq0.\end{align}
We note that the solution depends on the initial bubble through $v_{0}$
only in the overall normalization; the ratios $q_{\alpha}/q_{\beta}$
are independent of $v_{0}$.

Having determined the auxiliary quantities $U_{\alpha}(t)$, we can
now compute the non-Markovian volume distribution $V_{\alpha}(t)$
as\begin{align}
V_{\alpha}(t)= & \int_{0}^{t}\! dt_{0}U_{\alpha}(t_{0})e^{3H_{\alpha}(t-t_{0})}S_{\alpha}(t-t_{0})\nonumber \\
 & +\delta_{\alpha\alpha_{0}}e^{3H_{\alpha_{0}}t}S_{\alpha_{0}}(t).\label{eq:V alpha final 1}\end{align}
For $\alpha\neq0$, the integral in Eq.~(\ref{eq:V alpha final 1})
is dominated by $t_{0}\approx t$, which yields a term $\propto e^{3H_{0}t}$,
so the second term in Eq.~(\ref{eq:V alpha final 1}) is negligible.
Hence, by setting $Q(t)\approx R_{0}(t)$ and $q_{\alpha}=v_{0}\kappa_{0\rightarrow\alpha}$
one obtains the estimate\begin{equation}
V_{\alpha}(t)\approx\frac{v_{0}\kappa_{0\rightarrow\alpha}}{3H_{0}-3H_{\alpha}+\Gamma_{\alpha}}R_{0}(t)e^{3H_{0}t},\quad\alpha\neq0.\end{equation}
 We note that the ratios of volumes $V_{\alpha}(t)/V_{\beta}(t)$
are independent of the initial condition parameter $v_{0}$ and of
time, indicating a {}``stationarity'' of the solutions $V_{\alpha}(t)$
with $\alpha\neq0$. Moreover, these ratios are equal to the ratios
obtained in the Markovian regime, \begin{equation}
\lim_{t\rightarrow\infty}\frac{V_{\alpha}(t)}{V_{\beta}(t)}\approx\frac{c_{\alpha}}{c_{\beta}},\quad\alpha,\beta\neq0.\label{eq:V equal}\end{equation}
The imprecision in the above equality is exponentially small, of order
$O(\Gamma_{0}T)e^{-\Gamma_{0}T}$, as noted before. (To simplify calculations,
we also carried an imprecision of order $\kappa_{\alpha\rightarrow\beta}/(H_0-H_1)$
in the expressions for $c_{\alpha}$, but Eq.~(\ref{eq:V equal})
also carries that imprecision. 
This limitation is due to the approximations adopted in the present paper.)

It remains to compute $V_{0}(t)$. For $\alpha=0$, the estimation
of the integral in Eq.~(\ref{eq:V alpha final 1}) proceeds similarly
to the argument leading to Eq.~(\ref{eq:W 0 estimate}), except that
$R_{0}(t)$ is replaced by $S_{0}(t)$ which decays slower. The result
is\begin{equation}
V_{0}(t)\approx e^{3H_{0}t}\left[v_{0}S_{0}(t)+q_{0}\Gamma_{0}^{-1}Q(t)\right].\end{equation}
Since at large $t$ \begin{equation}
Q(t)\approx R_{0}(t)=S_{0}(t)\frac{\Gamma_{0\rightarrow\alpha}(t)}{\kappa_{0\rightarrow\alpha}}\ll S_{0}(t),\end{equation}
the dominant asymptotic for $V_{0}(t)$ for $t\gg T$ is \begin{equation}
V_{0}(t)\approx e^{3H_{0}t}v_{0}S_{0}(t).\end{equation}

\subsection{Discussion}

We will now interpret the results of the calculation in the previous
section. Since $Q(t)\ll S_{0}(t)$ at late times, the volume $V_{0}(t)$
within bubbles of type $0$ grows asymptotically faster than all other
$V_{\alpha}(t)$ for $\alpha\neq0$,\begin{equation}
\lim_{t\rightarrow\infty}\frac{V_{0}(t)}{V_{\alpha}(t)}\propto\lim_{t\rightarrow\infty}\frac{S_{0}(t)}{R_{0}(t)}=\infty,\quad\alpha\neq0.\end{equation}
 This indicates that the 3-volume at time $t$ is entirely dominated
by the bubbles of type 0, which we have labeled as those having the
largest Hubble rate $H_{0}=\max_{\alpha}H_{\alpha}$. Moreover, since
the integral in Eq.~(\ref{eq:V alpha final 1}) for $\alpha=0$ is
dominated by $t_{0}\approx0$, it follows that almost all of the volume
in bubbles of type 0 at time $t$ is in the \emph{very old} regions
of type 0. These regions of type 0 either belong to the original bubble
(if $\alpha_{0}=0$), or were nucleated early on (if $\alpha_{0}\neq0$)
and, by chance, have remained without decay for almost all of the
time $t$. This dominance does not depend on the initial conditions
and is due to the fact that non-exponential decay makes the nucleation
of other types of bubbles less likely in very old regions. The absolute
dominance of bubbles of type 0 will set in after time $T$. This is
different from the Markovian situation %
\footnote{The 3-volume is not a gauge-invariant quantity, and statements about
dominance of 3-volume at fixed time depend sensitively on the choice
of the time variable \cite{Winitzki:2005ya}. In particular, in Markovian
models the 3-volume is \emph{not} dominated by fastest-expanding bubbles
if one chooses the $e$-folding time $\tau\equiv\ln a$ as the time
coordinate. A similar gauge dependence is expected in the non-Markovian
case. The present calculation focuses on the effects of non-Markovian
decay, which are arguably more pronounced in the proper time gauge.%
} where bubbles of type 0 dominate with a \emph{finite} (but very large)
ratio, \begin{equation}
\lim_{t\rightarrow\infty}\frac{V_{0}^{\text{Markov}}(t)}{V_{\alpha}^{\text{Markov}}(t)}=\frac{c_{0}}{c_{\alpha}}\approx\frac{3H_{0}-3H_{\alpha}}{\kappa_{0\rightarrow\alpha}}\gg1.\end{equation}
Thus the qualitative picture of the distribution of volume in space
has changed due to the non-Markovian decay, but the change is not
drastic. This conclusion is similar in spirit to that obtained in
Sec.~\ref{sec:Non-Markovian-Sierpi}, where the fractal dimension
of the eternally inflating domain was modified from $3-\varepsilon$,
where $\varepsilon\ll1$, to exactly 3.

On the other hand, the 3-volumes $V_{\alpha}(t)$ within other types
of bubbles $\alpha\neq0$ grow proportionally to each other, and the
ratios $V_{\alpha}/V_{\beta}$ are almost the same (up to exponentially
small corrections) as those obtained in a Markovian calculation. Therefore,
any measure prescription that depends on the asymptotic ratios of
volumes, $V_{\alpha}/V_{\beta}$, will give unchanged predictions
as long as one asks about the volumes of bubbles of subdominant types
($\alpha\neq0$). Since the bubbles of type 0 (presumably, with a
Planck-scale $H_{0}$) are not especially interesting observationally,
one can conclude that a possible non-Markovian decay has no effect
on predictions obtained via any measure prescriptions based on volume
ratios.

The considerations in the present paper are limited to proper time gauge and to landscape scenarios satisfying the assumptions~(\ref{eq:assumption on 
kappa}). The methods developed here are applicable to landscapes of any type, and future work will show whether the conclusions hold in more general 
cases.

\subsection{Comoving distributions}

I now turn to considering the comoving distribution. One can define
the distribution $f_{\alpha}(t_{0},t)dt_{0}$ as the fraction of comoving
volume at time $t$ in bubbles of type $\alpha$ that were formed
at an earlier time between $t_{0}$ and $t_{0}+dt_{0}$. As before,
we set $f_{\alpha}(t_{0},t)=0$ for $t_{0}>t$; the volume remaining
from the initial bubble is not included in $f_{\alpha}(t_{0},t)$
but accounted for separately as the function $f_{\alpha_{0}}^{(0)}(t)$.
The formalism and the equations for the distribution $f_{\alpha}(t_{0},t)$
are quite similar to those developed above for the volume distribution
$V_{\alpha}(t_{0},t)$ except for the absence of the volume growth
factors $H_{\alpha}$. 

Instead of writing out the equations and the solutions for $f_{\alpha}(t_{0},t)$,
a simple consideration suffices to show that non-Markovian effects
are irrelevant for the distributions of comoving volume. The comoving
volume fractions $f_{\alpha}(t)$, defined regardless of age, exponentially
quickly become constant because the total comoving volume is conserved,
and the dominant eigenvalue of the relevant Markovian matrix is equal
to zero. The nucleation of bubbles will be always dominated by new
bubbles rather than by {}``aged'' comoving volume, simply because
the comoving fraction of the aged volume quickly goes to zero. In
fact, the {}``aged'' comoving volume has a smaller nucleation rate,
$\Gamma_{\alpha\rightarrow\beta}(t)\rightarrow0$ for $t\rightarrow\infty$,
and therefore plays an even less significant role in nucleation of
new bubbles as in Markovian models. This is in contrast to the situation
with the volume-weighted distributions, where the aged volume is rewarded
by an exponentially large extra growth factor $e^{3H_{0}t}$ compared
with the new volume that grows slower, as $e^{(3H_{0}-\Gamma_{0})t}$.
Therefore, the non-Markovian decay law will introduce only a vanishingly
small correction to the predictions obtained through comoving-volume
measure prescriptions.

\section{Is age-dependent decay cosmologically relevant?}

A subexponential asymptotic at late times is a generic feature of
quantum-mechanical systems. This feature can be understood heuristically
as follows~\cite{Krauss:2007rx}. Decay is due to the spreading of
the wave function away from the initial metastable state. However,
the wave packet keeps spreading even after tunneling out of the initial
domain. If the evolution proceeds without any wavefunction collapse
due to measurements, the tail of the outgoing wave packet will reach
back to the initial state. Since the spreading is a power-law process
(the root mean square uncertainty in position grows proportionally
to time), there will be a power-law tail of the wave packet that overlaps
with the initial domain. Hence, the probability of remaining in the
initial state has a power-law late-time asymptotic. These considerations
apply to tunneling processes in field theory as well because tunneling
occurs essentially along a one-dimensional path in field space, corresponding
to the instanton solution.

On the technical level, a necessary condition for the existence of
the subexponential asymptotic is that the Hamiltonian of the system
must be bounded (either from below or from above). An elementary consideration
is as follows. The probability of remaining in the metastable state
$\left|\psi\right\rangle $ is\begin{equation}
P(t)=\left|\left\langle \psi\right|e^{\textrm{i}\hat{H}t}\left|\psi\right\rangle \right|^{2},\end{equation}
where $\hat{H}$ is the total Hamiltonian of the system. Let us assume
that the spectrum of $\hat{H}$ is bounded from below, say by $E=E_{0}$.
Using the spectral decomposition, \begin{equation}
\hat{H}=\negmedspace\int_{E_{0}}^{\infty}\negmedspace E\hat{P}_{E}dE,\end{equation}
where $\hat{P}_{E}$ is an orthogonal projector onto the subspace
of energy $E$, we find\begin{equation}
\left\langle \psi\right|e^{\textrm{i}\hat{H}t}\left|\psi\right\rangle =\negmedspace\int_{E_{0}}^{\infty}\negmedspace e^{\textrm{i}Et}\left\langle \psi\right|\hat{P}_{E}\left|\psi\right\rangle dE\equiv\negmedspace\int_{-\infty}^{\infty}\negmedspace e^{\textrm{i}Et}\rho(E)dE,\end{equation}
where, by definition, the function $\rho(E)$ identically vanishes
for $E<E_{0}$. Because of the nonanalyticity of $\rho(E)$ at $E=E_{0}$,
the Fourier transform of $\rho(E)$ necessarily has a power-law asymptotic
$\propto t^{-d}$ at $t\rightarrow\infty$, where the power $d$ is
determined by the order of the (upper) nonzero derivative of $\rho(E)$
at $E=E_{0}$.

I conclude with some general comments regarding the plausibility of
the age-dependent decay in cosmological landscape scenarios. The subexponential
asymptotic was obtained by a quantum-mechanical consideration without
regard for gravitational effects. However, gravitation plays a central
role in vacuum decay~\cite{Coleman:1980aw}. Since the assumption
of a bounded Hamiltonian is important, while the Hamiltonian for General
Relativity is unbounded, it is not immediately clear that the subexponential
late-time decay will be manifest also when the effects of gravity
is taken into account. 

Another relevant consideration is the influence of measurements and
decoherence on the vacuum decay. The power-law asymptotic of the survival
probability holds only if the evolution of the wave function of the
metastable system is unitary and proceeds according to the Schrdinger
equation. The power-law decay can occur only if no wave function collapse
takes place during that evolution. Therefore, a direct observation
of the power-law decay is possible only if the metastable system as
well as any decay products are perfectly isolated and do not have
any possibility of interacting with any environment at least until
times $t\sim T$. It is clear that such a perfect and long-lasting
isolation is impossible in practice. Any realistic metastable system
and its decay products will interact with an environment long before
the crossover time $T$. After an interaction, the wave function will
effectively collapse back to the initial metastable state, and the
effects of the slow spreading of the wave packet will be removed. 

However, one needs to be careful when applying quantum-mechanical
considerations in the cosmological context. Since the potential observers
of vacuum decay are inside the decaying field configuration, it is
unclear whether they are able to effect a collapse of the wave function
of the entire Hubble patch around them. Several points of view are
possible. One could assume that a {}``measurement'' of the field
in the false vacuum state already occurs if sufficiently many gravitationally
interacting macroscopic bodies are present. In that case, the wave
function of the decaying field is continuously collapsing back to
the false vacuum configuration, and so it would appear that all vacuum
decay is \emph{entirely} inhibited due to the quantum Zeno effect
(QZE), whereby a metastable system does not collapse when continuously
measured. This conclusion appears implausible. On the other hand,
it is hard to implement a measurement of the field values on cosmological
super-horizon scales by any causal system. Hence, one could assume
that {}``measurements'' are absent until a tunneling event is completed
and a causally autonomous Hubble-size bubble of true vacuum is formed.
Then one finds that the late-time decay asymptotic is indeed relevant
to describing the landscape dynamics. Alternatively, one can suppose
that {}``measurements'' due to gravitationally induced decoherence
are effectively {}``performed'' only on super-Hubble time and distance
scales, as is the case in the decoherence of primordial quantum fluctuations
in an inflationary universe~\cite{Kiefer:1998jk,Kiefer:1998pb,Kiefer:1998qe,Kiefer:2006je}.
In this case, the QZE sets in only if the Hubble time is smaller than
the time scale of onset of the exponential decay law. In principle,
the QZE time scale can be estimated in a particular model of vacuum
decay. 

Presently, I merely summarized possible viewpoints on the relevance
of decoherence, the quantum Zeno effect, and subexponential decay
to cosmological evolution of false vacuum. More work is needed to
clarify this fundamental issue.

\section{Acknowledgments}

The author thanks Thomas Dent and Arthur Hebecker for helpful discussions.

\bibliographystyle{apsrev}
\bibliography{EI2}

\end{document}